\def\paperversion{cameraIEEE}

\def\paperversiondraft{draft}

\def\paperversioncameraIEEE{cameraIEEE}

\ifx\paperversion\paperversiondraft
\else
  \ifx\paperversion\paperversioncameraIEEE
  \else
    \def\paperversion{normal}
  \fi
\fi

\def\grammarlyon{on}

\ifx\grammarly\grammarlyon
\def\ClassReview{}
\else
\def\ClassReview{review,}
\fi

\ifx\paperversion\paperversioncameraIEEE
  \documentclass[conference]{IEEEtran}
\else
  \documentclass[\ClassReview anonymous, sigplan]{acmart}
\fi

\def\acmversionjournal{journal}

\ifx\paperversion\paperversioncameraIEEE
    \def\acmversion{none}
\else
  \makeatletter
  \if@ACM@anonymous
    \def\acmversion{anonymous}
  \else
    \def\acmversion{journal}
  \fi
  \makeatother
\fi

\usepackage{xcolor}

\PassOptionsToPackage{table}{xcolor}

\usepackage[utf8]{inputenc}
\usepackage[T1]{fontenc}
\usepackage{microtype}

\usepackage{xargs}
\usepackage{lipsum}
\usepackage[textsize=tiny]{todonotes}
\usepackage{xparse}
\usepackage{xifthen, xstring}
\usepackage[normalem]{ulem}
\usepackage{xspace}
\usepackage{marginnote}

\usepackage[frozencache=true,cachedir=minted-cache]{minted}
\usepackage{hyperref}
\usepackage{mathpartir}
\usepackage{dirtytalk}
\usepackage[ligature, inference]{semantic}
\usepackage{float} 
\usepackage{array}
\usepackage{colortbl}

\usepackage{listings}
\usepackage{balance} 
\usepackage{multicol}

\definecolor{codegreen}{rgb}{0,0.6,0}
\definecolor{codegray}{rgb}{0.5,0.5,0.5}
\definecolor{codepurple}{rgb}{0.58,0,0.82}
\definecolor{backcolour}{rgb}{0.95,0.95,0.92}

\lstdefinestyle{mystyle}{
  commentstyle=\color{gray},
  keywordstyle=\bfseries,
  numberstyle=\ttfamily\footnotesize\color{codegray},
  basicstyle=\linespread{0.8}\ttfamily,
  breakatwhitespace=false,
  breaklines=true,
  captionpos=b,
  keepspaces=true,
  columns=flexible,
  numbersep=5pt,
  showspaces=false,
  showstringspaces=false,
  showtabs=false,
  tabsize=2
}

\makeatletter
\font\uwavefont=lasyb10 scaled 652

\newcommand\colorwave[1][blue]{\bgroup\markoverwith{\lower3\p@\hbox{\uwavefont\textcolor{#1}{\char58}}}\ULon}
\makeatother

\ifx\paperversion\paperversiondraft
\newcommand\createtodoauthor[2]{%
\def\tmpdefault{emptystring}
\expandafter\newcommand\csname #1\endcsname[2][\tmpdefault]{\def\tmp{##1}\ifthenelse{\equal{\tmp}{\tmpdefault}}
   {\todo[linecolor=#2,backgroundcolor=#2,bordercolor=#2]{\textbf{#1:} ##2}}
   {\ifthenelse{\equal{##2}{}}{\colorwave[#2]{##1}\xspace}{\todo[linecolor=#2,backgroundcolor=#2,bordercolor=#2]{\textbf{#1:} ##2}\colorwave[#2]{##1}}}}
\expandafter\newcommand\csname #1f\endcsname[2][\tmpdefault]{
	\smash{\marginnote{
		\todo[inline,linecolor=#2,backgroundcolor=#2,bordercolor=#2]{\textbf{#1 (Figure):} ##2}}}
   }
}
\else
\newcommand\createtodoauthor[2]{%
\expandafter\newcommand\csname #1\endcsname[2][]{##1}%
\expandafter\newcommand\csname #1f\endcsname[2][]{##1}%
}%
\fi

\makeatletter
\patchcmd{\@addmarginpar}{\ifodd\c@page}{\ifodd\c@page\@tempcnta\m@ne}{}{}
\makeatother
\ifx\paperversion\paperversiondraft
  \makeatletter
  \if@ACM@journal
    \paperwidth=\dimexpr \paperwidth + 3.5cm\relax
    \oddsidemargin=\dimexpr\oddsidemargin + 0cm\relax
    \evensidemargin=\dimexpr\evensidemargin + 0cm\relax
    \marginparwidth=\dimexpr \marginparwidth + 3cm\relax
    \makeatletter
    \long\def\@mn@@@marginnote[#1]#2[#3]{%
      \begingroup
        \ifmmode\mn@strut\let\@tempa\mn@vadjust\else
          \if@inlabel\leavevmode\fi
          \ifhmode\mn@strut\let\@tempa\mn@vadjust\else\let\@tempa\mn@vlap\fi
        \fi
        \@tempa{%
          \vbox to\z@{%
            \vss
            \@mn@margintest
            \if@reversemargin\if@tempswa
                \@tempswafalse
              \else
                \@tempswatrue
            \fi\fi
              \rlap{%
                \if@mn@verbose
                  \PackageInfo{marginnote}{xpos seems to be \@mn@currxpos}%
                \fi
                \begingroup
                  \ifx\@mn@currxpos\relax\else\ifx\@mn@currxpos\@empty\else
                      \kern-\dimexpr\@mn@currxpos\relax
                  \fi\fi
                  \ifx\@mn@currpage\relax
                    \let\@mn@currpage\@ne
                  \fi
                  \if@twoside\ifodd\@mn@currpage\relax
                      \kern\oddsidemargin
                    \else
                      \kern\evensidemargin
                    \fi
                  \else
                    \kern\oddsidemargin
                  \fi
                  \kern 1in
                \endgroup
                \kern\marginnotetextwidth\kern\marginparsep
                \vbox to\z@{\kern\marginnotevadjust\kern #3
                  \vbox to\z@{%
                    \hsize\marginparwidth
                    \linewidth\hsize
                    \kern-\parskip
                    \marginfont\raggedrightmarginnote\strut\hspace{\z@}%
                    \ignorespaces#2\endgraf
                    \vss}%
                  \vss}%
              }%
          }%
        }%
      \endgroup
    }
    \makeatother
  \else
    \paperwidth=\dimexpr \paperwidth + 6cm\relax
    \oddsidemargin=\dimexpr\oddsidemargin + 3cm\relax
    \evensidemargin=\dimexpr\evensidemargin + 3cm\relax
    \marginparwidth=\dimexpr \marginparwidth + 3cm\relax
  \fi
  \makeatother
\fi

\definecolor{pairedOneLightBlue}{HTML}{a6cee3}
\definecolor{pairedTwoDarkBlue}{HTML}{1f78b4}
\definecolor{pairedThreeLightGreen}{HTML}{b2df8a}
\definecolor{pairedFourDarkGreen}{HTML}{33a02c}
\definecolor{pairedFiveLightRed}{HTML}{fb9a99}
\definecolor{pairedSixDarkRed}{HTML}{e31a1c}

\createtodoauthor{grosser}{pairedOneLightBlue}
\createtodoauthor{sid}{pairedThreeLightGreen}
\createtodoauthor{anurudh}{pairedFourDarkGreen}
\createtodoauthor{arjun}{pairedFiveLightRed}
\createtodoauthor{fehr}{pairedSixDarkRed}

\graphicspath{{./images/}}

\makeatletter
\newcommand\requiredelimiter[2][########]{%
  \ifdefined#2%
    \def\@temp{\def#2#1}%
    \expandafter\@temp\expandafter{#2}%
  \else
    \@latex@error{\noexpand#2undefined}\@ehc
  \fi
}
\@onlypreamble\requiredelimiter
\makeatother

\newcommand\newdelimitedcommand[2]{
\expandafter\newcommand\csname #1\endcsname{#2}
\expandafter\requiredelimiter
\csname #1 \endcsname
}

\newdelimitedcommand{toolname}{Tool}

\ifx\paperversion\paperversioncameraIEEE
\else

\fi

\usepackage{tikz}
\usetikzlibrary{arrows}
\usetikzlibrary{shapes}
\DeclareRobustCommand\circled[1]{\tikz[baseline=(char.base)]{
            \node[shape=circle,fill=pairedOneLightBlue,inner sep=1pt] (char) {#1};}}

\usepackage{booktabs}   
\usepackage{subcaption} 
\usepackage{amsmath}
\usepackage{mathtools}
\usepackage{marvosym}

\usepackage{caption}
\newcommand{\val}{\texttt{value}}
\requiredelimiter{\val}

\newcommand{\unk}{\texttt{unk}}
\requiredelimiter{\unk}

\newcommand{\absent}{\texttt{absent}}
\requiredelimiter{\absent}

\newcommand{\lpjump}{\textcolor{pairedTwoDarkBlue}{\texttt{lp.jump}}}
\requiredelimiter{\lpjump}

\newcommand{\lpjoinpoint}{\textcolor{pairedTwoDarkBlue}{\texttt{lp.joinpoint}}}
\requiredelimiter{\lpjoinpoint}

\newcommand{\lpswitch}{\textcolor{pairedTwoDarkBlue}{\texttt{lp.switch}}}
\requiredelimiter{\lpswitch}

\newcommand{\lppap}{\textcolor{pairedTwoDarkBlue}{\texttt{lp.pap}}}
\requiredelimiter{\lppap}

\newcommand{\lppapextend}{\textcolor{pairedTwoDarkBlue}{\texttt{lp.papextend}}}
\requiredelimiter{\lppapextend}

\newcommand{\lpint}{\textcolor{pairedTwoDarkBlue}{\texttt{lp.int}}}
\requiredelimiter{\lpint}

\newcommand{\lpbigint}{\textcolor{pairedTwoDarkBlue}{\texttt{lp.bigint}}}
\requiredelimiter{\lpbigint}

\newcommand{\CONS}{\texttt{cons}}
\requiredelimiter{\CONS}

\newcommand{\lpthunkify}{\textcolor{pairedTwoDarkBlue}{\texttt{lp.thunkify}}}
\requiredelimiter{\lpthunkify}

\newcommand{\lp}{\textcolor{pairedTwoDarkBlue}{\texttt{lp}}}
\requiredelimiter{\lp}

\newcommand{\lpforce}{\textcolor{pairedTwoDarkBlue}{\texttt{lp.force}}}
\requiredelimiter{\lpforce}

\newcommand{\lpconstruct}{\textcolor{pairedTwoDarkBlue}{\texttt{lp.construct}}}
\requiredelimiter{\lpconstruct}

\newcommand{\lpgetlabel}{\textcolor{pairedTwoDarkBlue}{\texttt{lp.getlabel}}}
\requiredelimiter{\lpgetlabel}

\newcommand{\lpproject}{\textcolor{pairedTwoDarkBlue}{\texttt{lp.project}}}
\requiredelimiter{\lpproject}

\newcommand{\lpinc}{\textcolor{pairedTwoDarkBlue}{\texttt{lp.inc}}}
\requiredelimiter{\lpinc}

\newcommand{\lpreturn}{\textcolor{pairedTwoDarkBlue}{\texttt{lp.return}}}
\requiredelimiter{\lpreturn}

\newcommand{\lpdec}{\textcolor{pairedTwoDarkBlue}{\texttt{lp.dec}}}
\requiredelimiter{\lpdec}

\newcommand{\lpblock}{\textcolor{pairedTwoDarkBlue}{\texttt{lp.block}}}
\requiredelimiter{\lpblock}

\newcommand{\rgn}{\textcolor{pairedSixDarkRed}{\texttt{rgn}}}
\requiredelimiter{\rgn}

\newcommand{\rgnval}{\textcolor{pairedSixDarkRed}{\texttt{rgn.val}}}
\requiredelimiter{\rgnval}

\newcommand{\rgnrun}{\textcolor{pairedSixDarkRed}{\texttt{rgn.run}}}
\requiredelimiter{\rgnrun}

\newcommand{\stdcall}{\texttt{std.call}}
\requiredelimiter{\stdcall}

\newcommand{\std}{\texttt{std}}
\requiredelimiter{\std}

\newcommand{\nofib}{\texttt{nofib}}
\requiredelimiter{\nofib}

\newcommand{\Int}{\texttt{Int}}
\requiredelimiter{\Int}

\newcommand{\mlir}{\texttt{MLIR}}
\requiredelimiter{\mlir}

\newcommand{\GHC}{\texttt{GHC}}
\requiredelimiter{\GHC}

\newcommand{\flrc}{\texttt{flrc}}
\requiredelimiter{\flrc}

\newcommand{\ghc}{\GHC}
\requiredelimiter{\ghc}
\newcommand{\Core}{\core}
\requiredelimiter{\Core}
\newcommand{\core}{\texttt{Core} }
\requiredelimiter{\core}

\newcommand{\EXPR}{\texttt{EXPR}}
\requiredelimiter{\EXPR}

\newcommand{\PHI}{\texttt{PHI}}
\requiredelimiter{\PHI}

\newcommand{\BB}{\texttt{BB}}
\requiredelimiter{\BB}

\newcommand{\BBID}{\texttt{BBID}}
\requiredelimiter{\BBID}

\newcommand{\INST}{\texttt{INST}}
\requiredelimiter{\INST}

\newcommand{\TERM}{\texttt{TERM}}
\requiredelimiter{\TERM}

\newcommand{\LABEL}{\texttt{LABEL}}
\requiredelimiter{\LABEL}

\newcommand{\FLAG}{\texttt{FLAG}}
\requiredelimiter{\FLAG}

\newcommand{\ASSIGN}{\texttt{ASSIGN}}
\requiredelimiter{\ASSIGN}

\newcommand{\VID}{\texttt{VID}}
\requiredelimiter{\VID}

\newcommand{\lambdapure}{\ensuremath{\lambda}\texttt{pure}}
\requiredelimiter{\lambdapure}

\newcommand{\lambdarc}{\ensuremath{\lambda}\texttt{rc}}
\requiredelimiter{\lambdarc}

\newcommand{\assignarrow}{\xrightarrow{~\texttt{asgn}~}}
\requiredelimiter{\assignarrow}

\newcommand{\phiarrow}{\xrightarrow{phi}}
\requiredelimiter{\phiarrow}

\newcommand{\controlflowarrow}{\xrightarrow{~\texttt{ctrl}~}}
\requiredelimiter{\controlflowarrow}

\newcommand{\timergnspeedup}{\texttt{1.0x}}
\newcommand{\timespeedup}{\texttt{1.09x}}

\colorlet{diffold}{red!80!black}
\colorlet{diffnew}{green!50!black}

\ifx\paperversion\paperversioncameraIEEE
\else
  \ifx\acmversion\acmversionjournal
  \acmJournal{PACMPL}
  \acmVolume{1}
  \acmNumber{1}
  \acmArticle{1}
  \acmYear{2017}
  \acmMonth{1}
  \acmDOI{10.1145/nnnnnnn.nnnnnnn}
  \startPage{1}
  \else
  \acmConference[PL'17]{ACM SIGPLAN Conference on Programming Languages}{January 01--03, 2017}{New York, NY, USA}
  \acmYear{2017}
  \acmISBN{978-x-xxxx-xxxx-x/YY/MM}
  \acmDOI{10.1145/nnnnnnn.nnnnnnn}
  \startPage{1}
  \fi
\fi

\ifx\paperversion\paperversioncameraIEEE
\else
\fi

\begin{document}

\title{Lambda the Ultimate SSA:\\Optimizing Functional Programs in SSA}                     



\ifx\paperversion\paperversioncameraIEEE
  \author{\IEEEauthorblockN{Siddharth Bhat}
  \IEEEauthorblockA{\textit{CSTAR} \\
  \textit{IIIT Hyderabad}\\
  Hyderabad, India \\
  siddharth.bhat@research.iiit.ac.in}
  \and
  \IEEEauthorblockN{Tobias Grosser}
  \IEEEauthorblockA{\textit{School of Informatics} \\
  \textit{University of Edinburgh}\\
  Edinburgh, United Kingdom \\
  tobias.grosser@ed.ac.uk}
}
\else
  \author{First1 Last1}
  \authornote{with author1 note}          
  \orcid{nnnn-nnnn-nnnn-nnnn}             
  \affiliation{
    \position{Position1}
    \department{Department1}              
    \institution{Institution1}            
    \streetaddress{Street1 Address1}
    \city{City1}
    \state{State1}
    \postcode{Post-Code1}
    \country{Country1}
  }
  \email{first1.last1@inst1.edu}          

  \author{First2 Last2}
  \authornote{with author2 note}          
  \orcid{nnnn-nnnn-nnnn-nnnn}             
  \affiliation{
    \position{Position2a}
    \department{Department2a}             
    \institution{Institution2a}           
    \streetaddress{Street2a Address2a}
    \city{City2a}
    \state{State2a}
    \postcode{Post-Code2a}
    \country{Country2a}
  }
  \email{first2.last2@inst2a.com}         
  \affiliation{
    \position{Position2b}
    \department{Department2b}             
    \institution{Institution2b}           
    \streetaddress{Street3b Address2b}
    \city{City2b}
    \state{State2b}
    \postcode{Post-Code2b}
    \country{Country2b}
  }
  \email{first2.last2@inst2b.org}         
\fi

\def\acmversion{none}

\ifx\grammarly\grammarlyon
\onecolumn
\else
\fi


\maketitle

\begin{abstract}

Static Single Assignment (SSA) is the workhorse of modern optimizing
compilers for imperative programming languages. However, functional languages
have been slow to adopt SSA and prefer to use intermediate representations based on minimal lambda calculi due to SSA's inability to express higher-order constructs.
We exploit a new SSA construct --- regions --- in order to express functional optimizations via classical SSA-based reasoning.
Region optimization currently relies on ad-hoc analyses
and transformations on imperative programs. These ad-hoc transformations are sufficient for imperative languages as regions are used in a limited fashion. In contrast, we use regions
pervasively to model sub-expressions in our functional IR.
This motivates us to systematize region optimizations. We
extend classical SSA reasoning to regions for functional-style analyses and transformations.
We implement a new SSA+regions based backend for LEAN4,
a theorem prover that implements a
purely functional, dependently typed programming language.
Our backend is feature-complete and handles all constructs of LEAN4's
functional intermediate representation \lambdarc{} within the SSA framework.
We evaluate our proposed region optimizations by optimizing \lambdarc{}
within an SSA+regions based framework implemented in MLIR and demonstrating
performance parity with the current LEAN4 backend.
We believe our work will pave the way for a unified optimization framework capable of representing,
analyzing, and optimizing both functional and imperative languages.
\end{abstract}

\begin{IEEEkeywords}
  Optimizing compilers, Functional programming
\end{IEEEkeywords}

\section{Introduction}

\begin{figure}
\includegraphics[width=\columnwidth]{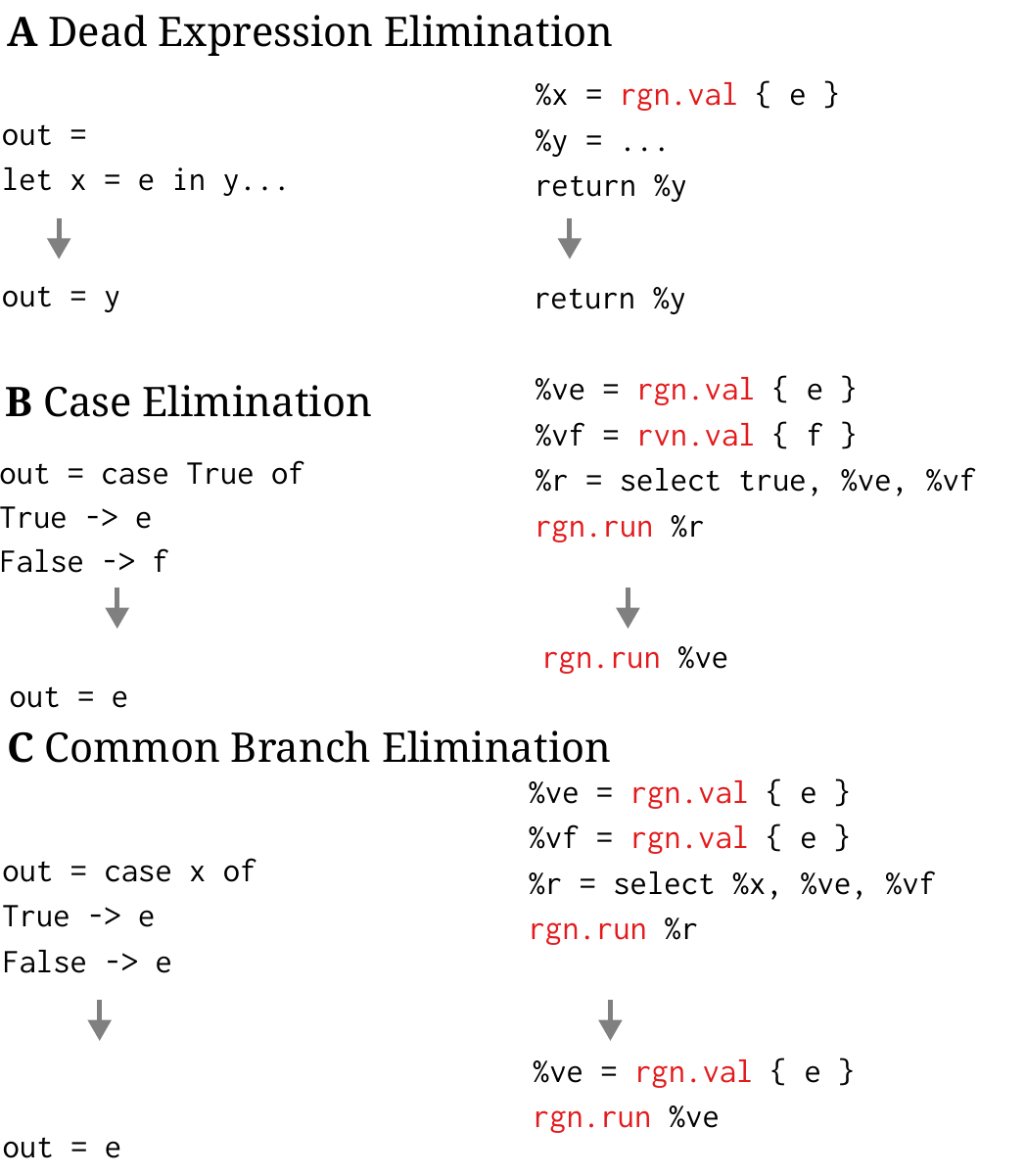}
\caption{We express regions as SSA values to adapt transformations in functional compilers
    in an SSA setting. We see that eliminating dead \texttt{let} bindings, eliminating \texttt{case}
    of known values, and fusing \texttt{case} branches are all variants of classical SSA transformations
    operating on region values declared by \rgnval{} and executed by \rgnrun{}.}
\label{fig:splash}
\end{figure}

Many optimizing compilers for imperative programming languages use Static Single Assignment (SSA) as their
intermediate representation (IR)~\cite{lattner2004llvm}\cite{Novillo2004DesignAI}. Such IRs impose structural and semantic rules on values to aid
the compiler's reasoning. Traditionally, such IRs express control flow using a Control Flow Graph (CFG) -- a flat collection of sequences of instructions (basic blocks) that
transfer control amongst one another. On the other hand, functional intermediate representations
use an \emph{expression-based} IR, where control flow is represented via particular expressions such as
case statements. An IR for functional constructs necessarily needs the ability to manipulate
sub-expressions.

The famous slogan ``SSA is continuations'' \cite{kelsey1995correspondence} is not entirely true as stated; the correspondence
between SSA and continuation-passing-style (CPS) is established between SSA and a
\emph{syntactically restricted} fragment of CPS, which is \emph{further annotated} with information
about which continuations represent intra-procedural control flow and which represent
inter-procedural control flow. Thus, the above translation is not practically useful in order
to design an SSA-based intermediate representation for functional programming languages, and
most functional IRs continue to use expression-based intermediate representations.
This has resulted in a schism within the compiler community, where the infrastructure built around
optimizing compilers for imperative languages does not get reused for functional languages
and vice versa.

In this work, we set out to heal this schism by providing \emph{convenient, easily analyzable and
optimizable} encodings of core functional IR constructs within SSA. Our key innovation is
to use regions to represent \emph{functional sub-expressions as SSA values}.
This allows us to extend SSA's core strength, the ability to reason about values via
sophisticated algorithms, to also cover reasoning about sub-expressions.
This enables us to mostly reuse SSA algorithms,
and in some cases extend SSA algorithms, to recover key functional
language optimizations within SSA. We also gain the ability to use the tooling
provided by MLIR's compiler infrastructure for (a) testing, (b) having a stable textual
and in-memory representation, (c) sophisticated
infrastructure for parallel peephole rewriting, and (d) support for parallelization and vectorization.

We contribute this novel encoding of a functional intermediate representation,
and an implementation of the encoding within the MLIR compiler framework. We also design
and implement new algorithms and modifications to existing
algorithms to reason about regions as sub-expressions. We test our design
by building a fully-functional backend for the LEAN4 \cite{moura2021lean}, functional programming language.
Our new backend uses the MLIR compiler framework to implement SSA+regions. We embed
LEAN4's intermediate representation, \lambdarc{}, within MLIR and show how to perform
common functional programming language optimizations on \lambdarc{} via SSA-based reasoning.
Concretely, our contributions are:
\begin{itemize}
    \item A minimal intermediate representation \lp{} which allows functional constructs
    to be lowered, analyzed, and optimized within an SSA setting.
  \item A novel use of regions to encode functional constructs within an imperative SSA-style IR, called \rgn{}.
  \item The design of functional-style optimizations for \rgn{}, which are analogues of
    classical SSA optimizations, applied to regions (\autoref{fig:splash}).
    \item An evaluation of our feature-complete backend for the LEAN4
      compiler and its \lambdarc{} intermediate representation which demonstrates
      the soundness of our approach towards representing functional programs in SSA via \lp{} and \rgn{}.
\end{itemize}

\section{Background}

\begin{figure*}[h]
\includegraphics[width=\textwidth]{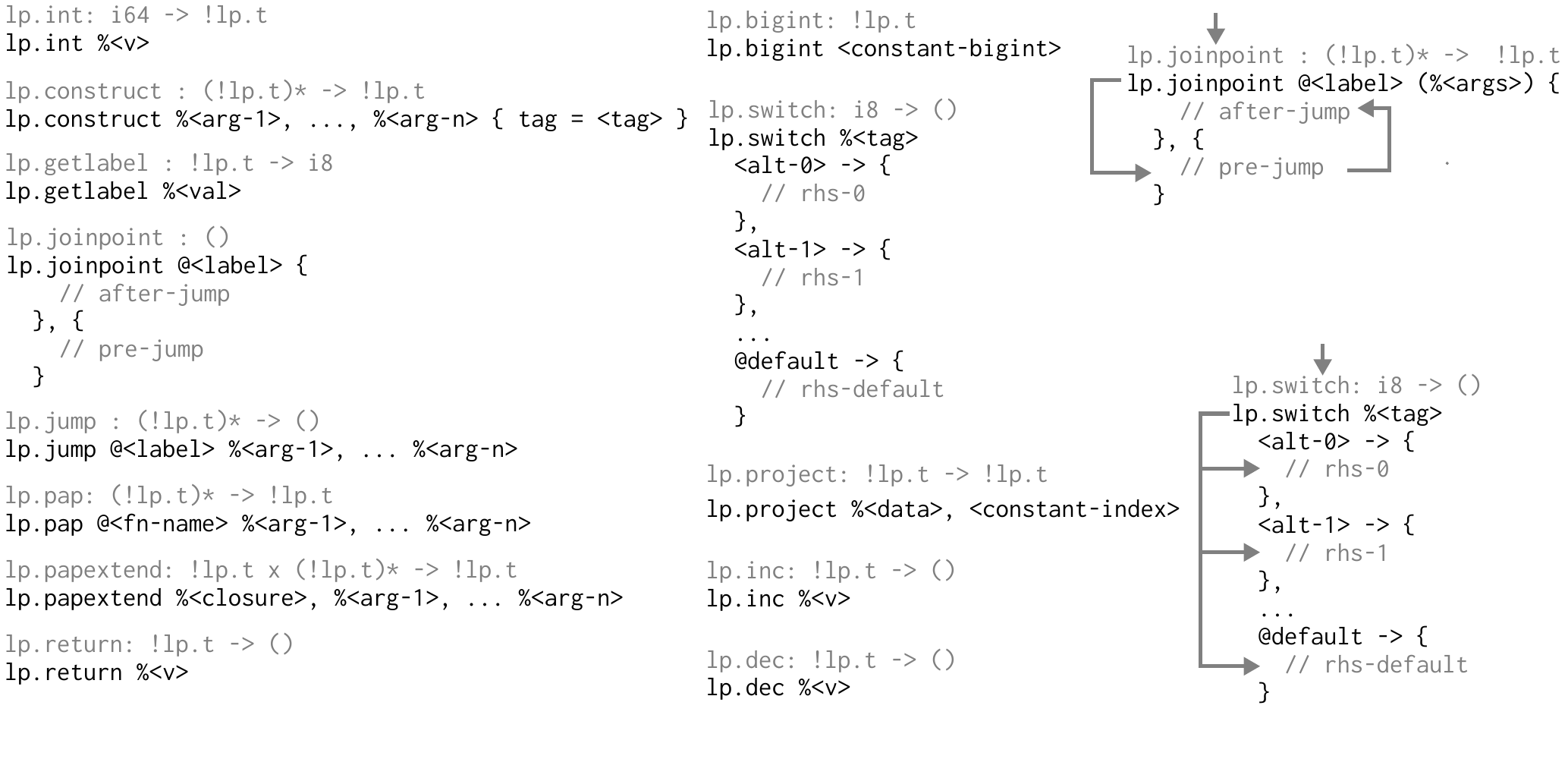}
\caption{\lp{}'s operations, their syntax and their types. Variable length argument lists are denoted by \texttt{(arg-ty)*}.
\lp{} expresses data constructors, pattern matching, closures,
and join-points within an SSA framework. The control flow of \lpjoinpoint{} and \lpswitch{} are shown on the right.
Control flow proceeds as per the arrows.}
\vspace{.7em}
\label{fig:syntax}
\end{figure*}

We provide background on SSA-based IRs, LEAN4, and functional programming
constructs.

\subsection{Static Single Assignment (SSA) and Regions}

An intermediate representation (IR) is in SSA form
\cite{rastello2010ssa} if each variable is assigned exactly once and no
variable remains undefined. SSA has gained popularity in imperative compilers
such as GCC~\cite{stallman2002gnu}, LLVM~\cite{lattner2004llvm}, and many
others, as data-flow information is explicitly expressed through dependencies
from the definition of a value to its uses (def-use chains). SSA-based IRs
typically use basic blocks that hold lists of sequentially executed operations,
each taking a list of argument values and returning a tuple of return values.
Terminator operations at the end of each basic block, which either branch to
another basic block or return from a function, combine these basic blocks into a
control flow graph (CFG). While IRs typically use a flat CFG,
MLIR~\cite{lattner2020mlir} recently introduced nested control flow as a
first-class concept to support abstractions that require control over scoping.
Operations can now receive regions, nested single-entry sub-CFGs, as additional
arguments. Regions make it easy to express concepts such as loop bodies,
branches of an if-statement, and case statements in functional programs.

\subsection{LEAN, \lambdapure{}, and \lambdarc{}}

LEAN\footnote{We refer to LEAN 4 when using the versionless expression LEAN.} \cite{de2015lean} is an open source theorem prover based
on a minimal dependently typed \cite{pierce2005advanced} kernel. After type checking
the LEAN program, the compiler compiles the program to \lambdapure{}, a
minimal, pure, strict, higher order intermediate representation that is suitable
for further lowering into assembly. \lambdapure{} is lowered to \lambdarc{}, an extension
of \lambdapure{} with reference counting. The current LEAN backend then lowers
\lambdarc{} to C.  Runtime features such as primitives for I/O, numerics,
reference counting, and task-based parallelism are provided by a custom runtime library,
\texttt{libleanrt}.



\subsection{MLIR}

MLIR \cite{lattner2020mlir} is a new compiler infrastructure that
aims to simplify the development of domain-specific compilers. For this, MLIR
provides a minimal SSA-based intermediate representation (IR), with the
ability to easily instantiate extensions to the core IR  (known as \emph{dialects})
which follow  SSA conventions. Having a customizable IR allows compiler
developers to model domain-specific concerns by introducing custom types,
operations, and attributes. Here, we briefly describe the relevant aspects of
MLIR that are used in our compiler.

\subsubsection{Modules}
Each program in MLIR is called a \emph{Module}. A Module consists of several
global functions. Function names such as \texttt{@foo} are global and allow
for linking function calls across modules. A function consists of basic-blocks. Each
basic-block is a sequence of operations, ending with a terminator operation.
SSA values are local to the scope of the function and have
names of the form \texttt{\%bar}.

\subsubsection{Operations}
SSA values are produced by \emph{Operations}, such as \texttt{addi}, which
stands for the integer addition operation. Each operation takes zero or more
SSA operands that are defined before it and returns zero or more SSA values.
Operations can also have compile time constants attached to them, such as
\texttt{\{phase = 90.0 : f64\}}. These are called \emph{attributes}.
\subsubsection{Dialects}
A \emph{Dialect} in MLIR is a collection of operations and types. The type
system in MLIR consists of either primitive types or user defined custom types to encode
more complex type systems. In this paper, we use two existing dialects, \texttt{std} and \texttt{scf}.
The \texttt{std} dialect contains all basic operations such as constant
declarations, arithmetic operations and memory manipulation. The Structured
Control Flow Dialect \texttt{scf} contains \texttt{if-else} and \texttt{for}
loop constructs.

\section{Expressing Functional Programs via SSA}
In this section, we survey \lambdarc{}, LEAN's intermediate representation
and discuss our embedding of \lambdarc{} into MLIR's SSA-based compiler IR.

\begin{figure}[H]
\includegraphics[width=0.5\textwidth]{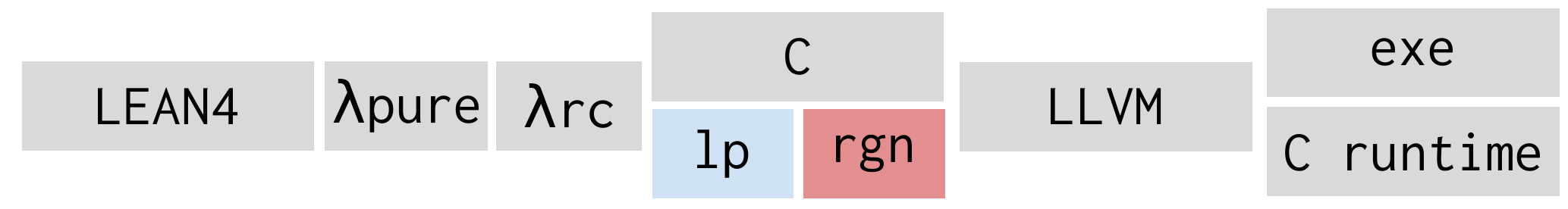}
\caption{LEAN4 compilation pipeline with our alternative \lp{} backend, supported by \rgn{} to encode functional constructs. We reuse
the LEAN frontend, as well as the LEAN runtime to ensure compatibility with the runtime and FFI.}
\label{fig:pipeline}
\end{figure}

LEAN4's \lambdapure{} IR is a minimal functional language with support for expressing
data constructors, pattern matching, function applications, and closure creation.
A lower level IR, known as \lambdarc{}, extends \lambdapure{} with operational concerns such as reference counting
within the same IR. The current LEAN4 compiler compiles \lambdarc{} down to C, followed
by invoking a C compiler to link with the runtime and generate binaries. Our MLIR backend
generates an MLIR dialect (\lp{}) from \lambdarc{}, and then continues compilation using the MLIR
compilation pipeline (\autoref{fig:pipeline}).

We introduce the \lp{} dialect (which stands for \lambdapure{}, though we also support \lambdarc{} reference counting instructions)
within MLIR, with the following set of high-level operations (\autoref{fig:syntax}):
\begin{itemize}
\item \lpint{}, \lpbigint{} to create machine integers and GMP-based big-integers.
\item \lpswitch{} to pattern-match on integers.
\item \lpconstruct{} to create algebraic data types.
\item \lpgetlabel{} to extract the tag of an algebraic data type.
\item \lpproject{} to extract out fields from an algebraic data type.
\item \lppap{}, \lppapextend{}, to express partial function applications (closure creation).
\item \lpjoinpoint{}, \lpjump{}, for representing structured control flow using join points.
\item \lpinc{}, \lpdec{} for reference counting.
\item \lpreturn{} to return values from \lp{} control flow.
\end{itemize}

We note that the dialect is \emph{feature complete} to represent a functional programming
language. It supports all the core constructs necessary, including closures and partial applications,
data constructors and pattern matching on data constructors, as well as support for reference counting.
This completeness is a strength of \lp{}, since we are able to compile a realistic functional
programming language (LEAN4) with our compiler pipeline. Like \lambdarc{}, the \lp{} dialect uses a
single type, denoted by \texttt{!lp.t}, to represent values that live on the heap, i.e., boxed values. We also use
standard integer types such as \texttt{i32} and \texttt{i8} as necessary to express interactions
with machine integers. \lp{} is \emph{type erased}, as we have erased most typing information from the
LEAN source program. We are left with just enough of the type information to generate code that is aware
of types such as \texttt{int}, \texttt{float}, pointers, as well as integer bit widths.

\begin{figure}[h]
\includegraphics[width=0.5\textwidth]{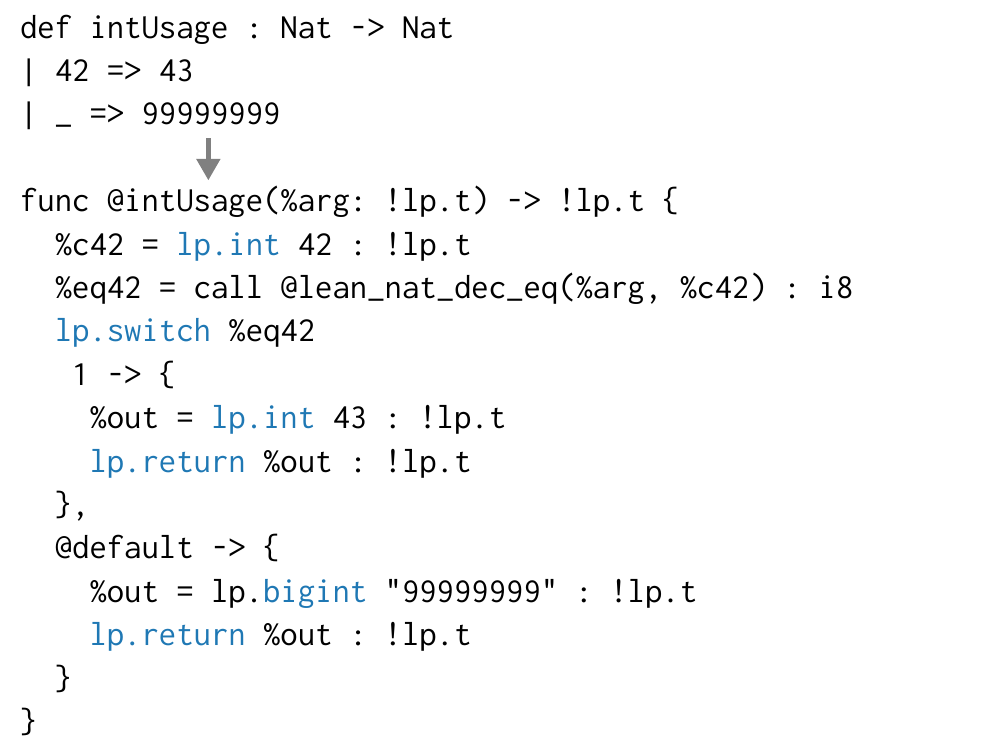}
\caption{Program demonstrating integer manipulation with
    \lpint{}, \lpbigint{}, and \lpswitch{}.
    We use \lpswitch{} to pattern match on raw integer values.
    LEAN integers are type-erased to create a uniform representation
    for small and large integers.
    Integers are compared using the
    runtime call \texttt{@lean\_nat\_dec\_eq}.}
\label{fig:caseint}
\end{figure}

\subsection{Integers \& Switch Cases}

\begin{figure}[h]
    \includegraphics[width=0.5\textwidth]{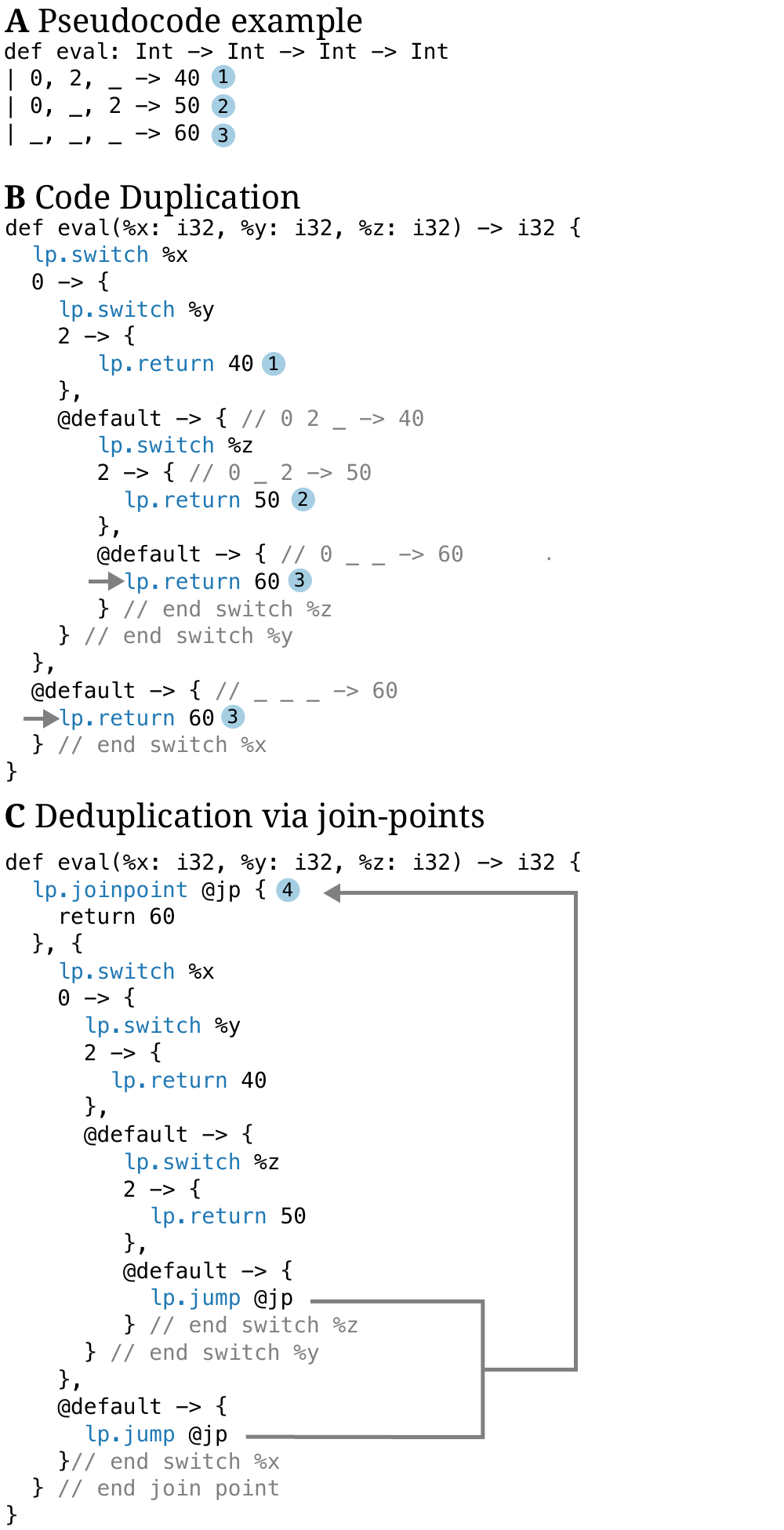}
    \caption{Complex pattern matching introduces code duplication across branches of control flow. This pattern matching is lowered using \lpjump{} to optimize the right hand side. The pattern matcher uses
        \lpjump{} to jump to the deduplicated right hand side. The arrows on the left indicate code duplication; the arrows on the right
        indicate deduplication via \lpjump{}.}
    \label{fig:lean-case-join-point-jp}
\end{figure}

LEAN supports arbitrarily sized natural numbers and integers with the \texttt{Nat} and \texttt{Int} types. LEAN
also guarantees that small integers are represented by a machine word. To support this,
\lp{} introduces the \lpbigint{} operation to construct arbitrary sized integers and
the \lpint{} operation to construct machine word sized integers. The LEAN4 compiler
lowers \lpbigint{} to runtime calls to create big-integers, while
\lpint{}s are lowered to machine integers.

Note that the runtime does not distinguish between naturals and integers --- they lower
to the same signed machine level representation,
and are only distinguished at the LEAN level
by the dependently typed theory.



As LEAN integers can be both machine integers and big-integers, pattern matching on a LEAN integer
must anticipate the scrutinee as being either a machine integer or a big-integer. \lambdarc{} hides this
complexity by staging the pattern match:
(1) \arjun[First, we  check whether one LEAN integer is equal to another]{It was not clear to me why we are comparing two integers until I reached the "for concreteness" paragraph.} using a runtime call to \texttt{@lean\_nat\_dec\_eq}.
This function handles equality checking between machine-machine, machine-bigint, and bigint-bigint integers uniformly.
(2) If the integers are equal, we execute the pattern match.

In more detail, the \texttt{@lean\_nat\_dec\_eq} function receives two integers of type
\texttt{!lp.t} as input, and returns either a \texttt{0} or a \texttt{1} of type \texttt{i8}
to indicate
whether the inputs are equal.
We switch-case on the return value with the \texttt{lp.switch} operation which dispatches control to a
matching switch arm. If no switch arm matches, control is dispatched to the \texttt{@default} arm
(\autoref{fig:syntax}).


For concreteness, consider the example program in \autoref{fig:caseint}. To compile the
first arm of the pattern match \texttt{42 =>}, we call \texttt{lean\_nat\_eq\_dec}
to check if the argument \texttt{\%arg} equals \texttt{42}. If this is true,
we proceed to execute the right hand side of the pattern match and return \texttt{43}.
The other arm \texttt{\_ =>} of the switch becomes the \texttt{@default} branch of the
\lpswitch{} which returns \texttt{99999999}.

\subsection{Join Points}

The naive lowering of case statements to switch operations can result in
code duplication across the right hand sides of nested cases.
We wish to avoid such duplication to ensure efficient compilation of pattern
matching. Consider a case statement that matches on two arguments
(\autoref{fig:lean-case-join-point-jp}). For the first two cases \circled{1}
and \circled{2}, we expect that the first argument is \texttt{0}, and the
second or the third argument is \texttt{2}. If this is not the case,
we execute the default case \circled{3}.

As \lpswitch{} can only match on a single integer at a time,
we introduce two nested \lpswitch{} statements
when lowering this pattern match.
The outer switch analyzes the first argument and the inner one analyzes the
second argument. If either case \circled{1} or case \circled{2} fails, we must
execute the code for the default case \circled{3}. When directly lowering
to nested switch operations, the expression that computes the result of the
default case is duplicated into multiple branches of the switch operations
(duplication indicated by arrows in \autoref{fig:lean-case-join-point-jp} \textbf{A}).
In case of complex pattern matches with many
variables, this duplication can be very costly.

To remedy such code duplication, as well as to increase the expressiveness of
the IR to encode complex control flow, LEAN uses \emph{join-points}
\cite{downen2016sequent, maurer2017compiling}. These join-points introduce the
ability to create \emph{labels} within the IR that control flow can jump to.
Intuitively, a label is a \emph{point} in the program, and jumping to a label
\emph{joins} control flow across the program at that point. The code duplication
that we observed in \autoref{fig:lean-case-join-point-jp} \textbf{A}
is eliminated via join-points (\autoref{fig:lean-case-join-point-jp} \textbf{B}). We create a
common region \circled{4} that performs the default action of returning \texttt{60}. Code duplication is
avoided  by introducing
two \lpjump{}s as indicated by the arrows.
Conceptually, the combination of \lpjoinpoint{} and \lpjump{} allows us
to represent local, named closures within the IR,
since the join-point region can refer to values that were defined before the join-point,
as well as the arguments that are passed by \lpjump{}. The difference between the join-point
closures and regular closures is that join-point closures are known to not escape, and all of
their call sites are within the function body. This allows join-points to be efficiently lowered
to jumps, unlike regular closures which require us to build heap objects that represent the closure.
Together, \lpjoinpoint{} and \lpjump{} provide a powerful tool to decrease code
duplication incurred by complex pattern matching.



\grosser{
    (2) Can we discuss alternative choices of how to reduce code duplication in the IR, e.g. the use
    of multi-dimensional switch statements. Are there any (dis)advantages in favour of each?
    (RESPONSE 2) I am hesitant to say things without being aware of all the trade-offs. The way I see it,
    you eventually have to lower multi-dimensional switches as well. I would imagine multidim-switch would lower
    into some kind of jump-table. I'd have to think about how it differs. One advantage of join-points
    is that they're literally gotos, so GHC uses them to also represent loops. (I am unsure whether LEAN4 does this.
    I think it does, I wouldn't bet my life on it).
}


\subsection{Data Constructors \& Pattern Matching}
In this section, we describe \lp{}'s support for constructing
data types via data constructors, and destructuring data types
via pattern matching. A data constructor is conceptually a
tagged union. Thus, we construct data using \lpconstruct{}, which
receives a tag and a sequence of arguments. The tag denotes which variant of a given data type is being built, and the argument list denotes the fields of the data constructor.

\begin{figure}[h]
\includegraphics[width=0.5\textwidth]{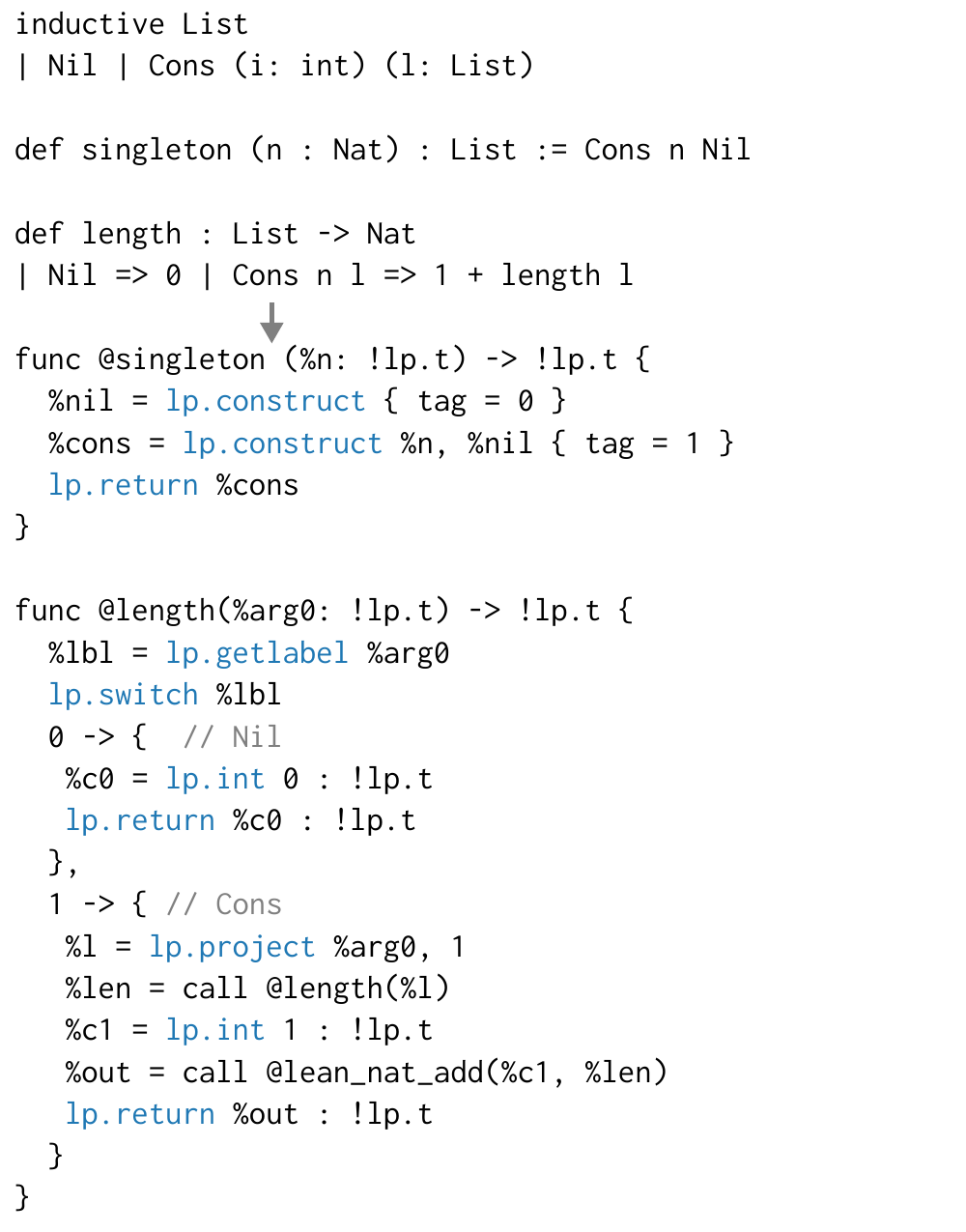}
\caption{We encode a data constructor with \lpconstruct{},
which receives the tag and the values held by the constructor as arguments.
We encode a case expression as a combination of \lpgetlabel{}, \lpproject{}, and \lpswitch{}.
\lpgetlabel{} extracts the label of the data constructor
that is switched on by \lpswitch{}. The fields of the value that is being \texttt{case}'d upon are extracted using \lpproject{}.}
\label{fig:constructor-and-case}
\end{figure}

For example, if we wish to build linked lists (\autoref{fig:constructor-and-case}), we would need two tags,
one for the empty list (\texttt{Nil} / tag \texttt{0}) and one
for a cons-cell (\texttt{Cons} / tag \texttt{1}). The function
\texttt{@singleton} constructs a list with a single cons-cell by
constructing a nil value and linking a cons-cell to the nil value.

To compute the length of the linked list, we pattern match on the
tag of the data constructor. If the tag is \texttt{0} (the object is
\texttt{Nil}), we return zero. Otherwise, the tag must be \texttt{1} (the object is \texttt{Cons}) and so we recursively call length on the
rest of the list and add one. We extract the tag via \lpgetlabel{},
pattern match on the tag using \lpswitch{}, and extract the list pointer
from the cons-cell using  \lpproject{}.

\begin{figure*}[h]
\includegraphics[width=\textwidth]{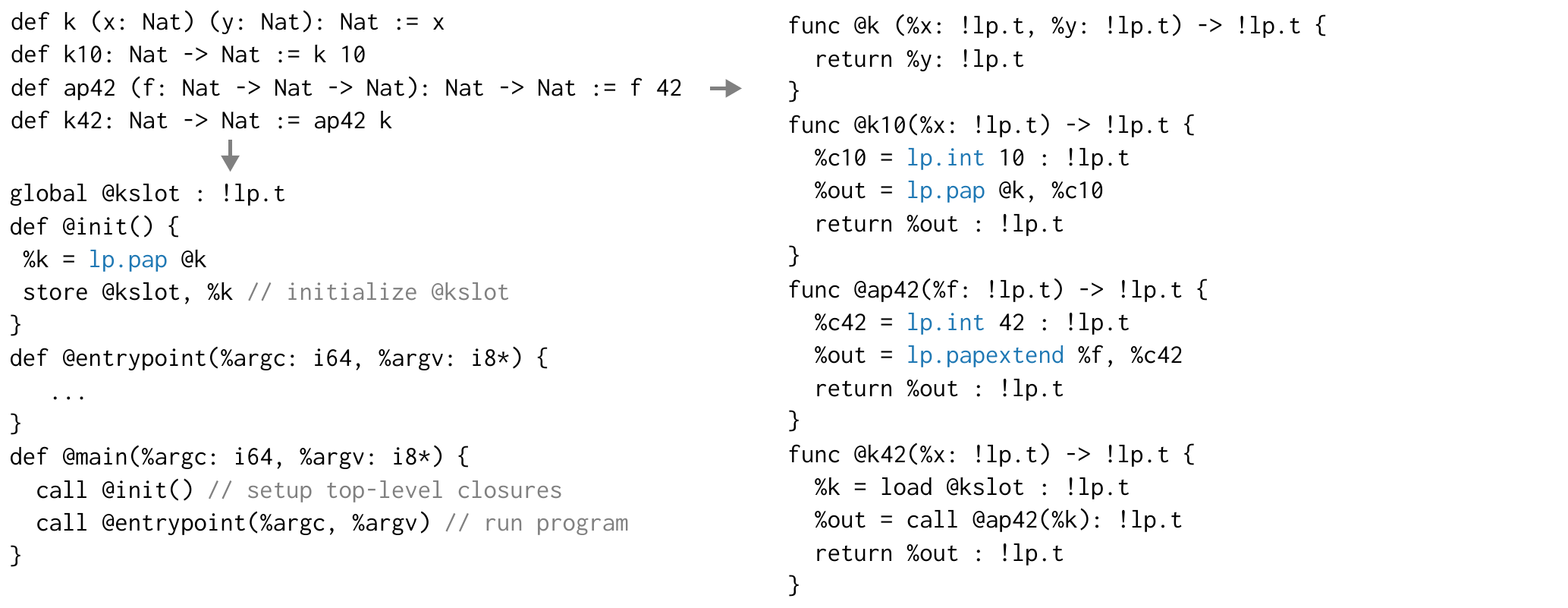}
\caption{Closures are built using \lppap{}, and closures are extended using \lppapextend{}. We see that
these are ubiquitous operations which almost always occur in the presence of higher order functions.
A program written in the functional style will typically have many closures, so a compact
and optimizable representation is important. \lambdarc{} performs lambda lifting which creates top-level closures such as \texttt{@kslot} which must be initialized before the program is run. We first run \texttt{init()} which initializes closures, followed by \texttt{entrypoint} which is the entry point to the LEAN program.}
\label{fig:lean-closures}
\end{figure*}

\subsection{Closures}

LEAN supports two types of function application: (1) regular (saturated) function applications where the caller
provides as many arguments as the arity of the function,
(2) partial function applications, where the caller provides fewer parameters than the function's arity.
In the case of a regular function application, we call the function eagerly, and reuse MLIR's function call infrastructure
to lower these. For partial
function applications, we build a \emph{closure} that holds onto the arguments that are supplied, and waits
for further arguments to be provided before the call can be made.
Once a closure receives all its arguments, the function held by
the closure is invoked with all the arguments that are held by the closure.


Partial applications require some care. First, to build a closure, we introduce the \lppap{} (partial application)
operation, which partially applies a function to some
arguments and builds a closure. To \emph{extend} a closure with more arguments, we introduce the
\lppapextend{} (partial application extend) operation, which takes as inputs a closure
and further arguments to extend the closure.  If these extra arguments
\emph{saturate} the closure (i.e., the closure now has all arguments), then
the function stored in the closure is invoked with arguments stored in the closure.
If the closure is not yet saturated the closure is \emph{extended} by storing the newly passed arguments in the closure.

Consider the example in \autoref{fig:lean-closures}. The function \texttt{@k10} partially applies \texttt{k} to the value \texttt{10}.
This is done by calling \lppap{} with arguments \texttt{@k} and \texttt{10}. This creates a closure which waits for the other argument \texttt{y} to invoke \texttt{k(10, y)}.

The function \texttt{@ap42} takes an arbitrary closure and applies it to the value \texttt{42}. This is achieved by using \lppapextend{} to extend
the closure of \texttt{f} with the argument \texttt{42}.

Finally, we draw attention to subtletly when invoking
\texttt{ap42} with the argument \texttt{k}. \texttt{ap42}, which obeys LEAN's
calling convention, expects a \emph{closure} as an argument.
But what is the closure associated
to \texttt{k}? We don't have one, as \texttt{k} is a top-level, raw
function, not a LEAN closure. This mismatch requires us to be able
to create \emph{empty closures} for top-level functions. This is
performed by the runtime, where we have an initialization phase
which creates top-level closures. In this case, the closure associated to \texttt{k} is initialized in the global variable
\texttt{@kslot} by the function \texttt{@init}.

As for lowering, \lppap{} and \lppapextend{} are lowered into runtime calls which manipulate closures.  A closure is represented in-memory by a function pointer and a list of pointers to the arguments which are held by the closure.

\subsection{Tail Calls}

\lambdarc{} explicitly keeps track of calls which must be tail calls for
memory consumption guarantees.  To respect this, we generate an LLVM call
annotated with a \texttt{musttail} attribute, which enforces that these calls
must be tail calls. If a call that is annotated as a \texttt{musttail} cannot be guaranteed to be
a tail call, then the LLVM optimizer provides an error. This is critical as tail-call-optimization is guaranteed by the LEAN language semantics, but is not guaranteed by the lowering to C as the C standard
does not require guaranteed tail call elimination.





\subsection{Reference Counting}

Since LEAN is a functional language, one does not explicitly refer to memory allocation
or de-allocation. Hence, LEAN implements automatic memory management
via reference counting. To incorporate reference counting for objects that
are created on the heap, we expose the operations \lpinc{} and
\lpdec{} to increment and decrement reference counts. These are lowered
to corresponding LEAN runtime calls.

\subsection{Linking against the Lean Runtime}

We link against the LEAN runtime which is written in C, by compiling to LLVM, and then linking using
\texttt{llvm-link}. We perform this step as many performance critical runtime routines, including refcounting,
are written in C, and are inlined by the \texttt{leanc} backend. Thus, for performance parity,
we compile the C code down to LLVM and link the runtime, thereby providing LLVM visibility to LEAN's runtime symbols.

\section{The \rgn{} Dialect}

In this section, we introduce the \rgn{} dialect to represent and optimize control flow
within \lambdarc{}. We describe the semantics of the \rgn{} dialect, the lowering from
\lp{}'s control-flow constructs (\lpswitch{}, \lpjoinpoint{}, \lpjump{}) to \rgn{} constructs,
and the adaptation of classical SSA optimizations to \rgn{}. The \rgn{} dialect
has two instructions:

\begin{itemize}
\item The \rgnval{} instruction creates an SSA value which names a region definition. These named values
 declare regions prior to being called. This is conceptually a \emph{continuation}, where the region represents
 a computation that is to be performed when invoked.
\item The \rgnrun{} terminator instruction transfers control flow to a region with the supplied arguments.
   This is conceptually \emph{invoking a continuation}. We branch to the region that is to be executed
    and continue execution from this region.
\end{itemize}

We allow \rgnval{} values to be passed as operands to MLIR's \texttt{select} and
\texttt{switch} instructions. These instructions allow us to pick a value based on
a boolean or an integer respectively. We use this to express choosing which
region is to be executed for control flow. We do not allow \rgnval{} operations
to interact with other operations; in particular, they may not
be passed to other functions as arguments and may not be returned.
This ensures that all uses of \rgnval{} can be statically analyzed
by our compiler, while still allowing us to leverage the analyses
and optimizations provided for \texttt{select} and \texttt{switch}.




\subsection{Lowering \lp{} to \rgn{}}
\begin{figure}[h]

    \includegraphics[width=\columnwidth]{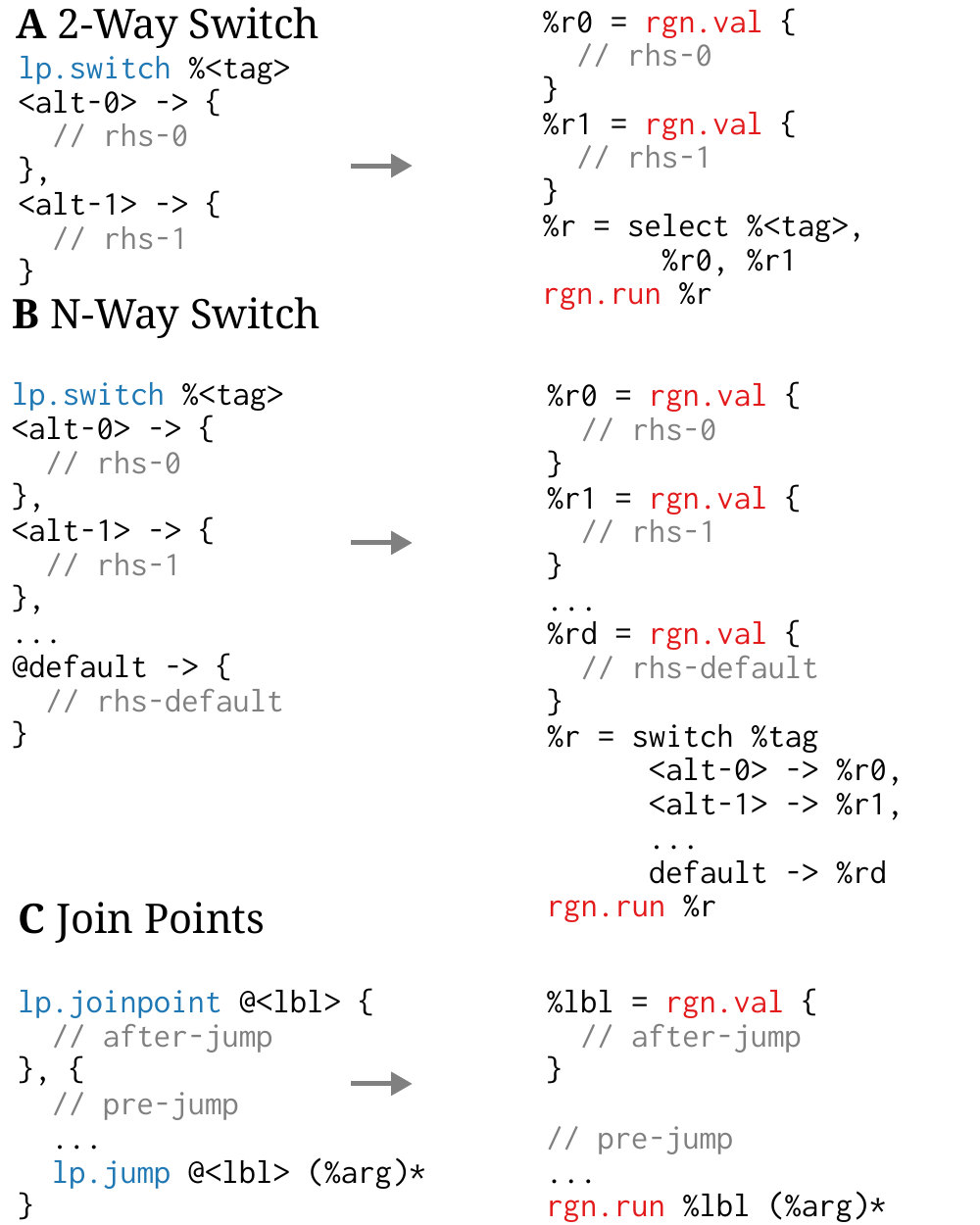}
    \caption{Lowering the control flow of \lp{} to \rgn{}. \lpswitch{} with two cases
       is lowered via \texttt{select} (\textbf{A}). \lpswitch{}
       with many cases is lowered via \texttt{switch} (\textbf{B}). \lpjoinpoint{} is lowered to a combination of \rgnval{} and \rgnrun{} (\textbf{C}).}
      \label{fig:lp-to-rgn}
\end{figure}

The lowering from \lp{}  to \rgn{} is straightforward (\autoref{fig:lp-to-rgn}). \lp{} has two control flow constructs,
\lpswitch{} \& \lpjoinpoint{}. \lpswitch{} is lowered by converting
every right hand side of a pattern match to a \rgnval{}, then selecting the correct right hand side that needs to be executed, and finally
executing the selected right hand side via \rgnrun{}. We lower the selection to either a \texttt{select} or
a \texttt{switch} depending on the number of cases. We lower \lpjoinpoint{} by converting the jump target to a \rgnval{}, the
\lpjump{} to a \rgnrun{}, replacing the joinpoint by the region that
is to be executed before the jump.

\subsection{Optimization Passes for \rgn{}}
In this section, we adapt classical optimization passes to \rgn{} 's
region-values. We explore optimizations that are made possible by these adapted
transformations.

\subsubsection{Dead Region Elimination}
Dead code elimination requires no changes to work with region values. If a region
value is never referenced, then it is never executed. It is thus dead and can
safely be removed.

\centerline{\includegraphics{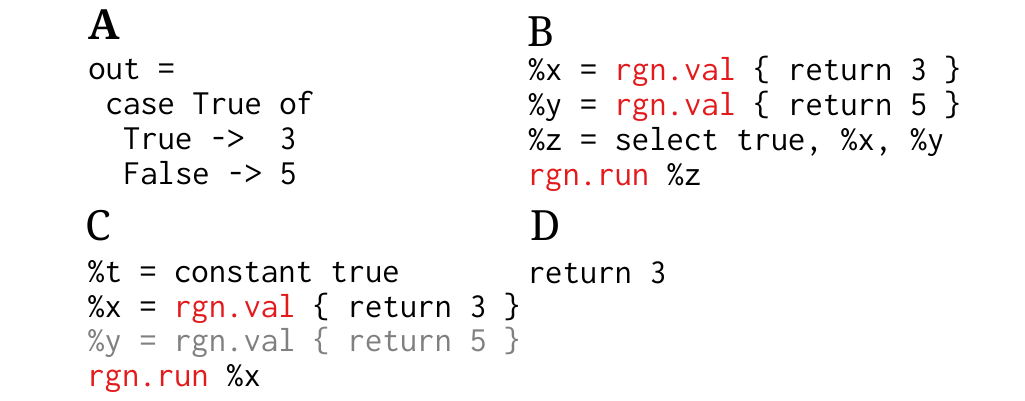}}

Figure \textbf{A} above is the original program. We show the naive translation to \rgn{} in \textbf{B}.
The optimizations for \texttt{select} on a constant value \texttt{true}
kick in and replace \texttt{\%z} by \texttt{\%x} in \textbf{C}. Finally, the
running of a known region is replaced by the constant value
that corresponds to the region in \textbf{D}.

\subsubsection{Global Region Numbering}

We introduce global value numbering for regions.\footnote{MLIR does not
perform global value numbering as it is unclear how to define value numbers for instructions with regions, as in general arbitrary regions do not yet have a prescribed semantics.}
For \emph{straight line regions}, the value number
of the region is defined as a rolling hash of the value numbers of all instructions within the region.
Two regions are defined to have the same value number if and only if the
sequence of instructions in the two regions have the same value numbers in
identical order. The restriction that a region has a single basic block is not
as restrictive as it first appears, since languages with high level control
flow can always be lowered to regions with a single basic block, with control flow expressed
via \rgn{} and select. \fehr{Maybe
bring back the lean example here, since it never uses basic blocks for
instance} Global value numbering of regions identifies redundant computations
across branches of control flow. This allows us to fold away equivalent
computations.

\sid{TODO: write down precise algorithm for region numbering}
\centerline{\includegraphics{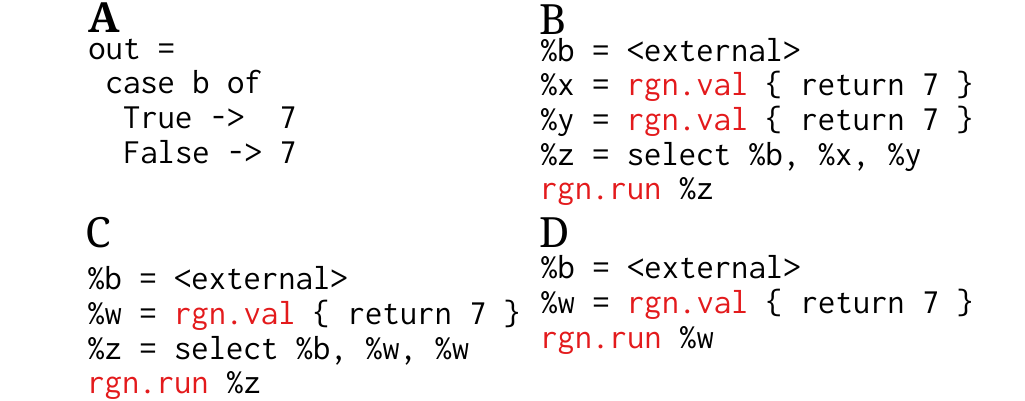}}

Figure \textbf{A} above is the original program. We show the naive translation to \rgn{} in \textbf{B}, where the value
\texttt{\%b} is external to the current scope being analyzed. The Region-CSE
algorithm fuses \texttt{\%x} and \texttt{\%y} into one region,
relabelled \texttt{\%w} in \textbf{C}. Finally, the select of two equal values
(\texttt{\%w}, \texttt{\%w}) on both branches is folded away in \textbf{D}.






\subsection{Lowering \rgn{} to \std{}}

In this section, we describe how to lower \rgn{} to a traditional SSA-based IR
without regions. For concreteness, we describe a lowering to MLIR's \texttt{std} dialect.
We anticipate no changes necessary to lower to another purely SSA based IR such as LLVM.

Since the semantics of \rgn{} is given entirely by adding extra structure
to flat CFGs, \rgn{} can be lowered by forgetting this extra structure.
The lowering is driven entirely by \rgnrun{}. First, we lower \rgnrun{}s by matching on the argument --- (1) A \rgnrun{} of a known \rgnval{}
is compiled to a branch of the region that is run, (2) A \rgnrun{} of a switch
(or select) is compiled to a jump-table. Finally, dead \rgnval{}
instructions are entirely dropped.







\section{Evaluation}

We evaluate our work by analyzing the correctness, performance, and memory
usage of the proposed SSA-based LEAN backend. Our evaluation was run
on an Intel Xeon CPU @ 2.20GHz with 126GB memory. We first
discuss the correctness of our compiler. We then consider the end-to-end
performance, by comparing runtimes of programs compiled by our compiler versus
the baseline Lean compiler on LEAN's benchmark suite, and then perform an
evaluation of the \rgn{} optimizations on the benchmark suite. Finally, we have
a holistic discussion about the costs and benefits of our new backend in
comparison to the current C backend of LEAN4.

\subsection{Correctness}

We test our compiler for correctness against the LEAN test suite, which
consists of 648 test cases, out of which we pass 648 (100\%) of tests.
This shows that our compiler correctly implements the semantics
of \lambdarc{} and interfaces properly with the LEAN runtime system.
This ensures that our evaluation is representative of functional
programming workloads.

\subsection{Performance Analysis}

We characterize the performance of our backend in comparison to LEAN's
default backend (commit hash \texttt{be4cf60}) using LEAN's benchmark
suite. The programs in the LEAN benchmark suite represent workloads commonly
encountered by functional programming languages:
\begin{itemize}
    \item \texttt{binarytrees} and \texttt{binarytrees-int}  implement a purely functional binary tree lookup, insert, and delete benchmark.
    \item \texttt{const\_fold}  implements constant folding on an expression based language.
    \item \texttt{deriv} benchmarks derivative computations on expression trees.
    \item \texttt{filter}  implements filtering values from a linked list based on a predicate.
    \item \texttt{qsort}  implements real in-place quicksort using LEAN's arrays.
    \item \texttt{rbmap\_checkpoint}  implements red-black tree insertion and lookup.
    \item \texttt{unionfind}  implements Tarjan's union-find algorithm.
\end{itemize}

\begin{figure}[h]
    \includegraphics[width=0.5\textwidth]{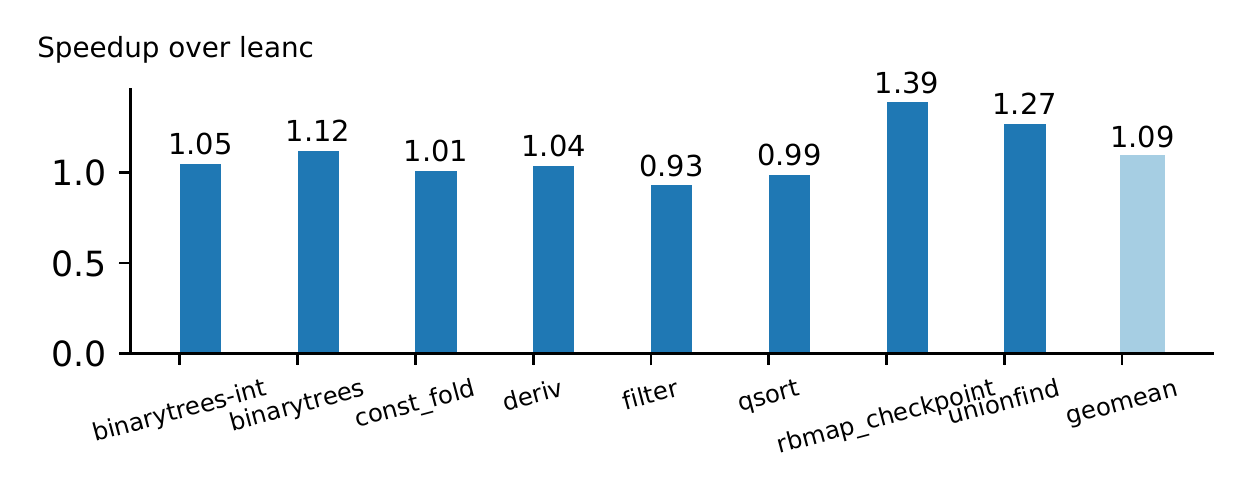}
	\grosserf{I prefer the geomean to be darker/highlighted, rather than the individual benchmarks. Also
	holds for the next figure}
    \caption{Speedup of our runtimes in comparison to LEAN4's existing C backend. The
        geomean speedup over the baseline LEAN4 compiler across all benchmarks is \timespeedup.
        Thus, we achieve performance parity with the LEAN4 backend.}
    \label{fig:speedup}
\end{figure}

The reported timing results (\autoref{fig:speedup}) are gathered as the
geometric mean of runtimes over ten runs. We report a geomean speedup of
\timespeedup~with the baseline LEAN compiler, which 
validates our claim that we achieve performance parity with the LEAN4 compiler.

\begin{figure}[h]

    \includegraphics[width=0.5\textwidth]{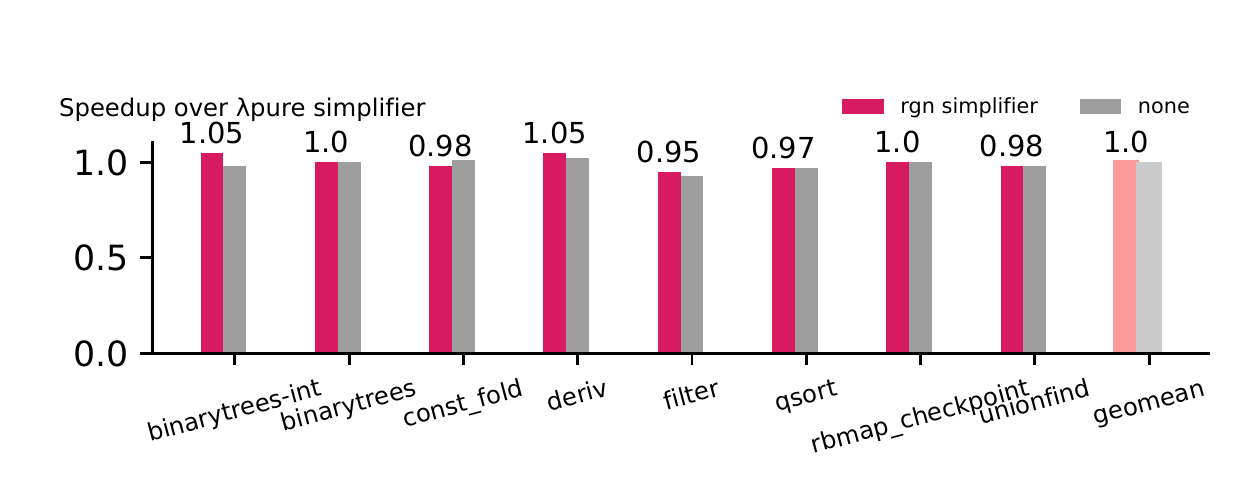}
    \caption{Speedup of \rgn{} dialect optimizations over \lambdarc{} (red) and speedup of no optimizations
    over \lambdarc{}  (gray). The numbers over the bar are the speedups of
         \rgn{} over \lambdarc{}. The geomean speedup of \rgn{} over \lambdarc{} across all benchmarks is \timergnspeedup.
         Thus, our simplification pipeline achieves performance parity with \lambdarc{}.}
    \label{fig:speedup-rgn}
\end{figure}

Finally, we compare the effectiveness of the \rgn{} dialect to that
of the \lambdarc{} simplifier (\autoref{fig:speedup-rgn}) by running three versions of the pipeline: (a) a baseline of our MLIR
pipeline which receives optimized code from the \lambdarc{} simplifier,
which we compare against (b) unoptimized \lambdarc{} code which is then
optimized by \rgn{} (we disable LEAN's  \texttt{simpcase} pass
which performs \rgn{} style switch simplification), as well as (c) unoptimized \lambdarc{} code
which is left unoptimized by \rgn{} before passing to LLVM. We find that the
performance is almost identical across the three variants. We conjecture
that LLVM's sophisticated control flow analysis is capable of
optimizing away the unoptimized IR we pass it.

In summary, we show that our SSA-based optimization pipeline matches the hand-written
LEAN optimization pipeline, while reasoning about functional constructs purely via SSA.
The \rgn{} dialect allows us to cleanly
express \lambdarc{}'s semantics with SSA, thereby making SSA+regions
an attractive choice to represent functional programming languages in traditional
imperative compiler tooling.

\subsection{Qualitative Analysis}
In this section, we compare and contrast our backend to LEAN's
backend from the perspective of a \emph{compiler engineer}. We first
survey a list of salient differences between LEAN's \lambdarc{} backend
and our MLIR-based backend (\autoref{fig:qualitative-analysis}).

\begin{figure}[h]
\begin{center}
\begin{tabular}[t]{b{1.2in}  b{0.8in} b{0.8in} }
\rowcolor{gray!30}
\emph{Feature} & \lambdarc{} + C & \lp{} + \rgn{} \\
Backend & C  & MLIR \\
\rowcolor{gray!10}
Vectorization & No & \texttt{affine}, \texttt{linalg} \\
Testing harness & \texttt{makefile} &  \texttt{FileCheck}, \texttt{llvm-lit}   \\
\rowcolor{gray!10}
Constant folding & Hand-written & MLIR rewriter \\
CSE & Hand-written & MLIR builtin \\
\rowcolor{gray!10}
DCE & Hand-written & MLIR builtin \\
Inliner & Hand-written & MLIR builtin \\
\rowcolor{gray!10}
Test minimization & None & \texttt{mlir-reduce} \\
Debug information & None & Possible  \\
\rowcolor{gray!10}
IDE support & None & LSP  \\
Tail call optimization & Heuristic & Guaranteed \\
\end{tabular}
\end{center}
\caption{Ecosystem differences between our MLIR-based backend and the current LEAN4 tooling.
Note that MLIR provides us access to a rich ecosystem of tooling for compiler development.}
\label{fig:qualitative-analysis}
\end{figure}

Notice the availability of rich tooling due to our use of the MLIR
compiler framework --- we are able to reuse existing optimization infrastructure
for constant folding, common subexpression elimination (extended by \rgn{}),
dead code elimination, and inlining. Similarly, we use LLVM's
test harness infrastructure for parallel test running and failure
reporting. The \lambdarc{} pipeline generates C code from which crashes
and miscompiles are difficult to debug as there is no association between the generated C program
and the source LEAN program.  One can potentially teach the LEAN frontend to preserve debug information
to be passed to MLIR. The MLIR  compiler framework makes retaining debug information
a top priority that enables the compiler engineer to correlate miscompiles
to LEAN program source locations accurately. Finally, as indicated before, LEAN requires
guaranteed tail call elimination which the current C based backend cannot. 
Our MLIR-based compiler guarantees this by using LLVM's
\texttt{musttail} annotations.

In summary, our compiler lays the foundation for LEAN to access an extensive
suite of analyses and optimizations. It also provides the compiler engineer tooling
for debugging and unit testing. Lastly, it allows us to provide deterministic
tail call elimination, which is necessary for the semantics of \lambdarc{}.

\section{Related Work}
In this section, we survey related work on functional intermediate representations
and uses of SSA in compilers for functional languages.

The Glasgow Haskell Compiler (GHC) \cite{jones1993glasgow} is an optimizing compiler for Haskell,
a lazy language with a focus on having an expressive type system and the ability to write
optimized programs. GHC uses an intermediate representation known as Core,
which is a strictly typed encoding of $F_\omega$ \cite{pierce2002types}.
Most optimizations to do with
exposing computation happen at this layer. After this stage, the encoding is
lowered to STG \cite{jones1992implementing},
a lower-level intermediate representation in
administrative normal form \cite{flanagan1993essence}.
Next, STG is lowered to \texttt{C--}\cite{jones1999c}, which is a
target-independent assembly language that supports garbage collection.
From this stage, GHC can either emit assembly or
generate LLVM. The impedance mismatch is synthesized during the translation from
STG to \texttt{C--}. \texttt{C--} is a traditional assembly IR that is well-suited
to borrow traditional optimizations from imperative compilers. Unfortunately, it
is precisely at this step that GHC chooses to simultaneously lower the encodings
of laziness and algebraic data types, making the assembly hard to optimize.


The Intel Haskell research compiler \cite{liu2013intel} is a whole-program optimizing
compiler for Haskell which focuses on vectorization and other program transformations for
performance. This compiler starts from GHC Core, and then
translates to a lazy ANF (Administrative Normal Form) \cite{flanagan1993essence}.
The compiler performs demand analysis and abstract simplification on ANF. The
demand analysis is performed using traditional abstract interpretation techniques.
Next, this demand information is used to interpret the program and perform
abstract simplification. It then compiles to an intermediate representation
called MIL, which is a loosely typed CFG based, SSA-like intermediate
representation.  They represent laziness as heap values, and manipulate the
heap. Thus, their representation and analysis of lazy values uses memory semantics
instead of value semantics.  MIL also does not have a notion of
nested regions. Therefore, MIL extends the traditional control flow controls
with finer-grained information, called as \texttt{cut} and
\texttt{interproc}. Finally, MIL is vectorized, and then lowered to
Intel's low-level IR, Pillar, which mirrors LLVM \cite{lattner2004llvm}.
Finally, assembly is generated from Pillar.

GRIN \cite{boquist1996grin} is an alternative monadic intermediate
representation for lazy and strict functional programming languages which
explicitly represents heap manipulation and case analysis. Due to the monadic
encoding, it has pointer semantics, not value semantics. GRIN therefore chooses
to run a sophisticated whole-program points-to analysis to resolve these
pointers for optimization. GRIN is well-suited for the whole-program paradigm,
while we focus on optimizing using local, per-module information without
incurring the penalty of a costly global analysis.

Thorin~\cite{Leia2015AGH} is a higher-order, functional
IR based on continuation-passing style. Thorin chooses to not use
explicit nesting, and uses a dependency graph instead. This has
the advantage of providing greater flexibility during compilation.
In contrast, we use explicit nesting based on regions, which is
easier to analyze and adapt into an SSA based framework.


The MLTon compiler \cite{weeks2006whole} is a whole program optimizing compiler
for Standard ML. MLTon uses a variant of SSA to encode algebraic data types and
case analysis. They perform whole program compilation, and use aggressive whole
program analyses reminiscent of GRIN to analyze and eliminate overhead.
Furthermore, their adaptations to SSA do not involve regions. They choose to
encode special terminator operations that represent control flow by case
analysis.
\section{Conclusion}

In this work, we have described \lp{}, which implements a new backend for the LEAN4
compiler within MLIR. To optimize a functional language within an SSA compiler
framework, we introduce \rgn{}, an intermediate representation
that is designed for region analysis and optimization. We adapt classical
functional language transformations to the SSA setting by using and
extending SSA algorithms to operate on \rgn{}. Finally, we implement our
suggestions and demonstrate that our new LEAN4 backend based on \lp{} and \rgn{}
passes all regression tests and achieves performance parity with the existing
LEAN4 compiler. This foundation is  a force multiplier, as we can now express the
LEAN4 semantics in a compiler framework that has been designed to be analyzed,
optimized, and offers mature support for performance and regression testing. We
envision \lp{} and \rgn{}
together acting as the bedrock for SSA-based optimizing compilers for many more functional
programming languages.

\ifx\paperversion\paperversioncameraIEEE
\else
\begin{acks}                            

  \grantsponsor{GS100000001}{National Science Foundation}{http://dx.doi.org/10.13039/100000001}
  under Grant
  No.~\grantnum{GS100000001}{nnnnnnn} and Grant
  No.~\grantnum{GS100000001}{mmmmmmm}.  Any opinions, findings, and
  conclusions or recommendations expressed in this material are those
  of the author and do not necessarily reflect the views of the
  National Science Foundation.
\end{acks}
\fi

\appendix
\section{Artifact Appendix}

\subsection{Abstract}

The artifact’s goal is to show how regions+SSA allows us to
create an MLIR-based backend for LEAN which achieves performance parity
with the LEAN4 compiler. The artifact consists of a docker container with accompanying
scripts to replicate figure  9, 10. The docker container
is the only piece needed to run all the experiments. Scripts to
generate the figures and the table come with the docker container.


\subsection{Artifact Check-List (Meta-Information)}

{\small
\begin{itemize}
  \item {\bf Program: }A custom LEAN4 backend based on the MLIR compiler
    toolchain, along with LEAN4's test suite for testing, and LEAN4's benchmark suite for
    performance analysis.
  \item {\bf Compilation: }A C++ compatible compiler to bootstrap LLVM/MLIR as well as LEAN4.
  \item {\bf Run-time environment: }Any operating system that supports Docker.
  \item {\bf Hardware: }Any x86 machine.
  \item {\bf Output: }PDF files replicating \autoref{fig:speedup} and \autoref{fig:speedup-rgn}, and a successfull
    run of the entire LEAN4 test suite.
  \item {\bf How much disk space required (approximately)?: }10GB.
  \item {\bf How much time is needed to prepare workflow (approximately)?: }2 hours.
  \item {\bf How much time is needed to complete experiments (approximately)?: }1 hour.
  \item {\bf Publicly available?: }Yes.
\end{itemize}

\subsection{Description}

\subsubsection{How Delivered}
The artifact is delivered as a Docker container
and is available at \url{http://doi.org/10.5281/zenodo.5786074}.

\subsubsection{Hardware Dependencies}
None.

\subsubsection{Software dependencies}
The docker image has dependences needed to compile MLIR, Lean4, and our
compiler tooling to run the test suite.

\subsection{Experiment Workflow}

Access the docker image \texttt{cgo22.docker} from
(\url{http://doi.org/10.5281/zenodo.5786074}), then run:

\begin{minted}[fontsize=\scriptsize]{text}
$ docker load -i cgo22.docker
$ docker run -it siddudruid/cgo21-v4
$ su nonroot # switch to non-root user
$ export PATH=/code/llvm-project/build/bin:$PATH
$ export PATH=/code/lean4/build/release/stage1/bin:$PATH
$ export PATH=/code/lz/build/bin:$PATH
$ export LEANLIB=/code/lean4/build/release/stage1/lib
$ export LD_LIBRARY_PATH=$LEANLIB:$LD_LIBRARY_PATH
$ cd /code/lean4/build/release && \
     make -j4 test
$ cd /code/lz/test/lambdapure/compile/bench && \
     ./speedup-time.py --data --plot --nruns 10
$ cd /code/lz/test/lambdapure/compile/bench && \
     ./speedup-rgn-time.py --data --plot --nruns 10
\end{minted}


Upon running \texttt{make -j4 test}, the test output is printed to \texttt{stdout}.
The scripts \texttt{speedup-time.py} and \texttt{speedup-rgn-time.py},
produce PDFs \texttt{speedup-time.pdf} and \texttt{speedup-rgn-time.pdf}
in the directory \texttt{/code/lz/test/lambdapure/compile/bench/}:

\begin{minted}[fontsize=\tiny]{text}
/code/lz/test/lambdapure/compile/bench/speedup-time.pdf
/code/lz/test/lambdapure/compile/bench/speedup-rgn-time.pdf
\end{minted}

To open the pdf file, keep the container running, and in another
shell instance, use the \texttt{docker cp}
command to copy files from within the container out to the host:

\begin{minted}[fontsize=\footnotesize]{text}
$ docker container ls # find   ID
$ docker cp <CONTAINERID>:<PATH/INSIDE/CONTAINER> \
            <PATH/OUTSIDE/CONTAINER>
\end{minted}
For more about \texttt{docker cp}, please see:
(\url{https://docs.docker.com/engine/reference/commandline/cp/})

%
%
%
%
%

\subsection{Evaluation and Expected Result}
On running the test suite with:

\begin{minted}[fontsize=\footnotesize]{text}
$ cd /code/lean4/build/release && make -j test
\end{minted}

We find that the output is:

\begin{minted}[fontsize=\footnotesize]{text}
100% tests passed, 0 tests failed out of 648
\end{minted}

See that we pass all tests. The test script that is run can be found at
\texttt{/code/lean4/test/common.sh} which can be seen to invoke our compiler
pipeline.

To generate performance plots, we run:

\begin{minted}[fontsize=\footnotesize]{text}
$ cd /code/lz/test/lambdapure/compile/bench && \
    ./speedup-time.py --data --plot --nruns 10
$ cd /code/lz/test/lambdapure/compile/bench && \
    ./speedup-rgn-time.py --data --plot --nruns 10
\end{minted}

For \texttt{speedup-time.pdf}, we expect a geomean speedup of \texttt{1.0x} with the
baseline \texttt{leanc}, and for \texttt{speedup-rgn-time.pdf}, we expect a geomean speedup of \texttt{1.0x} between region
optimizations and no region optimizations.

\subsection{Full Workflow Example}

\begin{minted}[fontsize=\tiny]{text}
# Grab docker image from http://doi.org/10.5281/zenodo.5786074
$ curl https://zenodo.org/record/5786074/files/cgo22.docker?download=1 > cgo22.docker
$ docker load -i cgo22.docker
$ docker run -it siddudruid/cgo21-v4
$ su nonroot # switch to non-root user
$ export PATH=/code/llvm-project/build/bin:$PATH
$ export PATH=/code/lean4/build/release/stage1/bin:$PATH
$ export PATH=/code/lz/build/bin:$PATH
$ export LEANLIB=/code/lean4/build/release/stage1/lib
$ export LD_LIBRARY_PATH=$LEANLIB:$LD_LIBRARY_PATH
$ cd /code/lean4/build/release/ && make -j test # NO test failures
100% tests passed, 0 tests failed out of 648

Total Test time (real) = 399.28 sec
[100%] Built target test

$ cd /code/lz/test/lambdapure/compile/bench && \
./speedup-time.py --data --plot --nruns 10

$ cd /code/lz/test/lambdapure/compile/bench && \
./speedup-rgn-time.py --data --plot --nruns 10
\end{minted}


\IEEEtriggeratref{12}
\bibliography{references}


\end{document}